**Visualizing the Evolution of Twitter (X.com) Conversations: A Comprehensive Methodology Applied to AI Training Discussions on ChatGPT**

*Nicole Jess[1] and Hasan Gokberk Bayhan[2]*

With the rise of social media platforms, especially X.com (formerly Twitter), there is a growing interest in understanding digital social networks and human digital interactions. This paper presents a comprehensive methodology for extracting, processing, and visually analyzing data from X.com, using a combination of Python and R packages, enhanced by our publicly accessible, customizable code. Our approach compiles a dynamic dataset that captures various interactions: replies, retweets, and mentions. To explore deeper insights, the data is subjected to sentiment analysis and keyword coding, indicating shifts in discourse over time. Our method is structured in three primary phases. Initially, R is employed for pulling data and the formation of social network datasets. Following this, the combination of Python and R is utilized for sentiment analysis and keyword coding, aiming to uncover the underlying emotional shifts and language transitions within topics of discussion. The final phase employs R to visualize the dynamic shifts within these social networks. These visualization tools highlight changes in user interactions and patterns of influence. For a practical demonstration, we analyzed conversations on X.com regarding the controversial proposal to halt AI development, focusing specifically on discussions about ChatGPT. By using keyword searches, leading voices in the debate were identified. Our analysis of sentiment and keywords revealed patterns in emotions and language, while visual tools illustrated the development of network connections and their influence. This study emphasizes the vital role of visual tools in understanding online social dynamics in the digital age.

[1]Dept. of Counseling, Educational Psychology & Special Ed., Michigan State University, East Lansing, MI, 48224, USA. Email: jessnico@msu.edu
[1]Google Scholar: https://scholar.google.com/citations?user=47csG6oAAAAJ&hl=en&oi=ao
[2]School of Planning, Design, and Construction, Michigan State Univ., East Lansing, MI, 48224, USA. Email: bayhanha@msu.edu - ORCID ID: 0000-0003-1207-5491 (Corresponding Author)
[2]Google Scholar: https://scholar.google.com/citations?user=70LnSxIAAAAJ&hl=en&oi=ao

## 1. Introduction

The rise of social media platforms, such as X (formerly known as Twitter), has revolutionized the way people communicate, interact, and disseminate information. The vast amounts of data generated by these platforms offer researchers an unparalleled opportunity to study social networks and human behavior (Tolsma and Spierings, 2024). In recent years, numerous studies have employed social network analysis (SNA) techniques to analyze Twitter data, focusing on various topics, such as information diffusion (Kim et al., 2018), community detection (Kim et al., 2013), influence analysis (Shi et al., 2019), sentiment analysis (El-Masri et al., 2021), language use by different groups (Evolvi, 2019), and identification of influential people (Anger and Kittl, 2011). The large-scale and open nature of these networks provide researchers with rich datasets, ideal for diverse analytical and visualization methodologies. Furthermore, these platforms host a wide array of discussions rich in meaning, ranging from social and political topics to advancements in technology (Nik-Bakht and El-Diraby, 2017). However, there is a gap in the body of knowledge regarding comprehensive methodological guidance on how to transform raw Twitter data into a format suitable for analysis and how to interpret such vast amounts of complex data. This paper aims to address this gap by providing a step-by-step approach for creating meaningful social network datasets from Twitter data and generating useful dynamic visualizations based on the datasets.

Our methodology focuses on three primary interaction types on Twitter: replies, retweets, and mentions. While Twitter interactions also include quotes, we opted to focus on replies, retweets, and mentions for this study. The nature of quotes, which often reframe the original content within a new narrative or context, often adds an additional layer of complexity to sentiment and keyword analysis. Specifically, quotes can alter the sentiment or emphasis of the original message, requiring more nuanced analytical approaches that were beyond the purview of our current methodology (Flamino et al., 2023). Additionally, including quotes could bias the visual representations of the networks, as they may disproportionately highlight or obscure certain connections or themes. We demonstrated how to extract, process and organize interactions into edgelists and nodelists, forming visually intuitive networks that are easy to understand and effectively display dense information without clutter. Our integrated approach to data extraction and visualization allows users to gain a deeper understanding of these dynamics, in a more streamlined way. By observing dynamic visualizations, researchers can explore how connections in networks form and dissolve over time and consider sentiment and language changes related to the topic with an immediate understanding of complex outcomes, enhancing the accessibility of research findings. Moreover, the open-source code provided in the methodology section could be adapted according to future research needs such as including node characteristics of follower count and verification status. We offer this methodological guidance to assist researchers in tailoring our approach for their unique research needs in social network analysis using Twitter data.

Furthermore, we demonstrated the adaptability of our method by automating the analysis of sentiment and keywords, and by creating dynamic visualizations to assess the differences between pre- and post-periods related to the letter aiming to halt the training of AI, specifically, the relatively powerful GPT-4 models, for six months. The letter highlighted the societal dangers of advanced AI and played a pivotal role in alerting the public about the threats posed by advanced AI models. Numerous prominent tech leaders endorsed this letter, titled "Pause Giant AI Experiments: An Open Letter" (NPR, 2023).

The paper is structured as follows. Section 2 presents the literature, and section 3 covers the methodology, data extraction, processing, organization, and integration of four main steps: Pulling Data, Creating Edgelist and Nodelist, Text Analysis, and Visualizations. Section 4 demonstrates the applicability of our methodology using an example dataset of Twitter interactions, including the brainstorming discussions to explore the relevant topic, and discusses potential extensions and adaptations of the proposed method for various research objectives, such as analyzing the language, distinct characteristics, sentiment, and identification of influential individuals in the network. Finally, Section 5 concludes the paper and highlights future research directions in the realm of social network analysis using Twitter data.

## 2. Literature Review

### 2.1 Data Extraction from Online Platforms for Social Network Analysis

Social network analysis (SNA) is a methodological framework used to study the structure and dynamics of social relations among actors, focusing on the relationships between individuals, groups, or organizations. SNA enables researchers to analyze patterns and properties of relationships to better understand social phenomena with qualitative and quantitative measures (Wasserman and Faust, 1994).

As social media's popularity grows, researchers increasingly use these platforms for social network data. X.com is a widely used source, with the primary method of data extraction being Application Programming Interfaces (APIs), which allow researchers to access public tweets and user information. These APIs facilitate the gathering of large amounts of data and have been extensively used in various studies (Tufekci, 2014). In addition to APIs, some researchers use web scraping to extract Twitter data. This technique involves parsing web page HTML to retrieve structured information. Though effective, web scraping is generally more complex and time-consuming than API-based methods (McCormick et al., 2017). Large-scale datasets collected by third parties have also been used in Twitter research (Weng et al., 2010), containing millions of tweets and user profiles for analyzing broad patterns and trends. However, accessing these datasets can be challenging due to their limited availability and privacy concerns.

One of the most popular and convenient R packages for data extraction from X.com is the 'rtweet' package (v1.1.0) (Kearney, 2019). This package provides a user-friendly interface to the Twitter API v2, allowing researchers to collect tweets and profile information efficiently. The package offers several functions for data collection, including searching for tweets by keywords, filtering by language, retrieving user profiles, and obtaining follower/friend lists. Additionally, 'rtweet' is the only actively maintained package, in contrast to others like academictwitteR (Mastroeni et al., 2023), which has faced limitations due to recent API changes. With 'rtweet', researchers can scrape node and edge data, or capturing profiles and their connections. This capability is essential for creating edgelists, which are data structures representing the connections between nodes in a social network, consisting of node pairs. Such structured formats are crucial for social network analysis, enabling the examination of various connection types to explore research questions and compute network metrics like centrality, clustering coefficients, and modularity (Borgatti et al., 2018). However, creating edgelists from Twitter data presents challenges, including the overwhelming volume and complexity of data, noisy content such as spam and bots, potential biases in data collection, and handling missing or incomplete data (González-Bailón et al., 2011). Researchers must address these challenges to ensure the reliability and validity of network metrics, as the quality and relevance of edgelists directly influence the interpretation and applicability of network metrics.

**2.2 Text Analysis in Online Platforms**

Sentiment analysis, or opinion mining, blends computational linguistics, text analysis, natural language processing, and biometrics to analyze people's sentiments, attitudes, and opinions towards various entities such as products, services, and organizations. This method excels in identifying, extracting, and quantifying subjective information and emotional states, with wide applications in marketing, customer service, and clinical medicine, often analyzing 'voice of the customer' data from reviews, surveys, and social media (Agarwal, 2011).

In sentiment analysis, various models have showcased their efficacy for X.com data. DeepMoji, developed by Felbo et al. (2017), uses emojis to predict sentiments, trained on a substantial dataset of 1.246 billion tweets. Shelar & Huang (2018) utilized NLTK to analyze sentiment polarity in charitable tweets, while Chandrasekaran and Hemanth (2022) explored TextBlob's diverse NLP capabilities, including part-of-speech tagging and sentiment analysis. Pellert et al. (2022) demonstrated the versatility of Hugging Face's Transformers, suitable for a range of text-based tasks, and Hutto and Gilbert (2014) employed VADER, a lexicon and rule-based tool optimized for social media contexts. Our study focuses on TextBlob and VADER, chosen for their proven effectiveness across a broad range of NLP tasks and their established prominence in sentiment analysis literature, highlighting their reliability and broad applicability. In choosing computational tools for sentiment analysis, Python's widespread use in the field influenced our preference for Python packages, which best met our analytic needs as supported by heuristic evaluations.

Beyond sentiment analysis, keyword coding is crucial for understanding discourse changes over time. This method systematically identifies key words or phrases to reveal thematic shifts in text, useful in dynamic environments like X.com influenced by current events and trends. Kulkarni et al. (2015) demonstrated a computational approach to detect significant linguistic shifts using techniques ranging from frequency-based statistics to time series analysis, with examples from X.com. Similarly, Burt and Reagans (2022) explored how network structure and the timing of messages, or 'pulse', lead to specialized jargon in teams. By analyzing keyword frequency and co-occurrence, researchers can track evolving discussions and gauge the impact of public messaging.

**2.3 Dynamic Network Visualization**

The visualization of social networks is essential for understanding their structure, evolution, and dynamics. Static visualizations can provide valuable insights, but dynamic visualizations are particularly useful for capturing the temporal aspects of network change. Various tools and techniques have been developed for the dynamic visualization of networks, including Pajek, Gephi, and R packages such as networkDynamic, ndtv, and tsna (Bender-deMoll and Morris, 2021; Bender-deMoll, 2022; Butts et al., 2022).

The NetworkDynamic, ndtv, and tsna packages are crucial for studying network dynamics, enabling the visualization of evolving temporal patterns. O'Hara et al. (2020) used these tools to analyze swine shipments in China, providing key insights for disease control. Hammami et al. (2022) examined French swine trading partnerships, while Camarillo-Ramirez et al. (2020) assessed software usability for dynamic network analysis. Carlson et al. (2021) applied these packages to study resilience strategies in fisheries during COVID-19. Despite their broad applicability, these tools are underutilized in analyzing social network platform interactions due to complexity of datasets.

## 3. Methodology

The twitter-sna repository (Jess et al., 2023), which is available on GitHub, comprises a collection of R Markdown scripts. These scripts execute R and Python code, providing an efficient framework for conducting social network analysis with Twitter data. These scripts are publicly available, enabling other researchers to replicate and build upon our work. Figure 1 illustrates the sequence in which the scripts should be run to generate the inputs necessary for subsequent scripts. This paper provides a walkthrough of the scripts for pulling Twitter data, creating social network data, sentiment analysis, keyword coding, dynamic visualizations, and path visualizations. While we outlined the general process in this paper, readers can refer to the repository for further details and code, as well as additional scripts and modeling options which are not covered in this paper. X.com and Twitter are used interchangeably throughout this paper.

During the preparation of this work the authors used ChatGPT version 4.0 in order to improve readability and language. After using this tool/service, the authors reviewed and edited the content as needed and take full responsibility for the content of the publication.

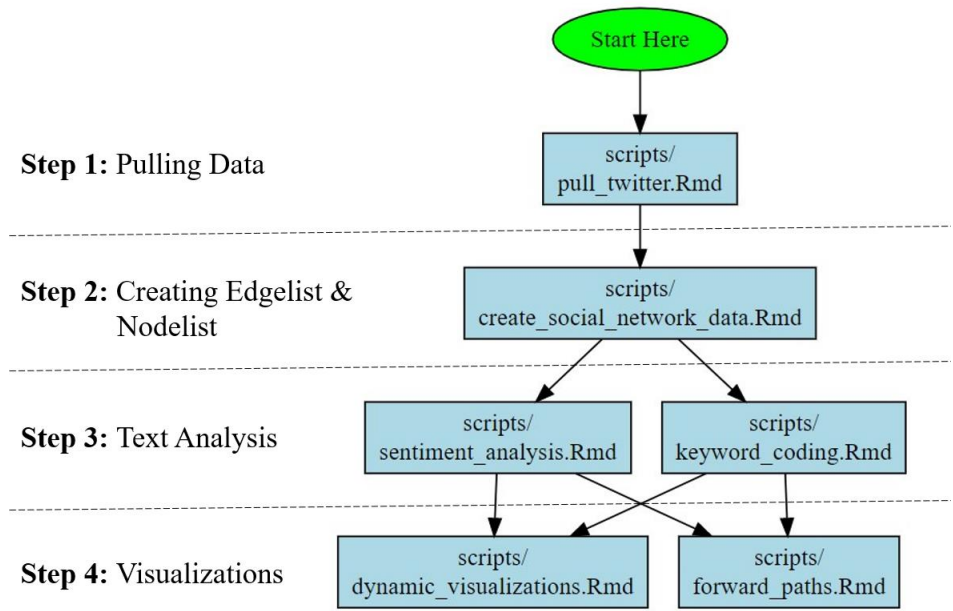

**Figure 1.** Flow diagram for methodology.

## 3.1. Pulling Twitter (X.com) Data (R)

Data pulling from Twitter denotes the automated procurement of publicly accessible information, including tweets and user details, facilitated by software libraries that interface with Twitter's API. The method for pulling Twitter data used in this paper utilized the rtweet (v1.1.0) package in R Statistical Software (v4.2.3) and academic research level access to Twitter's API. Since the time of pulling data for this paper, changes to the API have eliminated the free academic research access level so researchers will have to decide which level of paid access would be appropriate for the context of their research in terms of cost and the number and timing of the tweets relevant to the research. To use the following methods, researchers must have a Twitter Developer account which allows tweet lookup. This will provide a bearer token for making requests on behalf of the application and allow access to the information necessary for the procedures described in the following sections. The tools demonstrated in this paper could also be used with data from other social network platforms as long as the researcher is able to extract the text, date, and user interactions with posts related to the topic of interest.

For a Twitter query, researchers should select keywords related to their research topic and include any variations or abbreviations of those keywords that may be used on Twitter. In our case study examining attitudes towards ChatGPT before and after the AI training halt letter, we specifically included "large language model" and "LLM" as keywords. This ensured our queries targeted users familiar with these technical terms. A two-stage pulling strategy can often be useful for pulling a more manageable number of tweets about popular topics on Twitter. In stage one, researchers can use more specific keywords (e.g., "Elon Musk AI letter") to narrow the list of users to those who tweeted about that topic. In stage two, researchers can use broader keywords (e.g., "ChatGPT") but limit the pull to the users captured in stage one. This allows the researcher to extract the necessary data to explore how the conversation around the broader topic may relate to the narrower topic.

Researchers should also select an appropriate time interval for their query based on the timing of the Twitter conversation. Often, the time may be split into different intervals for different purposes. For example, researchers may consider dividing the time into pre and post-time intervals relative to some event that is of interest. Once the relevant keywords and time intervals have been selected, researchers can pull tweet data using the rtweet package's search_fullarchive() function.

## 3.2. Creating Edgelist & Nodelist for Social Network Data (R)

Twitter interactions, including mentions, replies, retweets, and quotes, can be used to construct the edges of the social network. Mentions are tweets that include another user's Twitter handle preceded by an "@" symbol, replies are tweets in which a user responds to another user's tweet, retweets are reposts of a tweet, and quotes are reposts of a tweet with a comment added. The choice of which types of interactions to use as edges depends on the research context. The code developed for this paper extracts the Twitter user ID data for users who interact with tweets pulled by the query and restructures the data into an edgelist format. The edgelist should include the following variables: 'tweet_id', which is a unique identifier for the tweet; 'created_at', which indicates the date and time the tweet was posted; 'from', which is the user that posted the tweet; 'to', which is the user who was mentioned, replied to, retweeted, or quoted by the user that posted the tweet; 'text', which is the full text of the tweet; and 'edge_type', which specifies the type of interaction (mention, reply, retweet, quote).

In addition to the edgelist, a corresponding nodelist should be created, which includes profile information for all the users involved in the edgelist. Twitter profiles provide users' names, descriptions, locations, verified status, follower counts, and other information. The nodelist is produced by generating a list of the unique user IDs found in the edgelist and pulling information from those Twitter user profiles.

### 3.3 Text Analysis

### 3.3.1 Sentiment Analysis (R and Python)

Our study uses text analysis to convert Twitter data into node and edge variables for exploring emotional aspects through visualization and modeling. Sentiment analysis, a refined text analysis technique, assesses emotional tones within the data, enabling a detailed examination of expressed sentiments.

Sentiment analysis can be conducted in Python using the TextBlob and VADER packages. Using the TextBlob package, the text of the tweets is passed to the constructor TextBlob(), which produces a sentiment.polarity attribute. Using the VADER package, the polarity_scores() method of the SentimentIntensityAnalyzer object is applied to the text of each tweet, which returns a 'compound' score, representing the overall sentiment polarity of the text. The polarity scores produced by these packages range from -1 (negative sentiment) to 1 (positive sentiment). In our case study, we used both packages to allow for a comparison of their respective performances.

### 3.3.2 Keyword Coding (R)

Keyword coding is a method that can be used to track the use of specific language in the context of a Twitter conversation. It's a valuable tool in text analysis, especially in examining large quantities of data. Researchers should generate a list of important key terms that they would like to track through the network. Using the grepl function in R, which searches for matches to the keyword argument in the text of the tweets, each tweet is assigned a "1" if it contains at least one keyword or a "0" if it does not contain any of the keywords. Keyword coding provides additional insights beyond sentiment analysis, allowing researchers to understand better the prevalence of particular words and how the use of those words can spread through the network.

## 3.4 Visualizations (R)

### 3.4.1 Dynamic Visualizations

Dynamic visualizations provide an essential tool for capturing and illuminating temporal variations in social networks, thereby enabling a comprehensive understanding of these networks' evolution and interaction dynamics over time. The dynamic visualizations described here were generated using three R packages: networkDynamic, ndtv, and tsna. These packages are designed for use with dynamic networks, which are social networks that change over time.

Dynamic networks require an onset and terminus for each edge to specify when the connection starts and ends, respectively. In our case study, the onset was set to the date the tweet was posted, and the terminus was set to four days after posting. The selection of interval duration should be carefully determined, considering the rate of change, the context of the conversation, and the interaction's popularity, aligned with the researcher's objective. The data must then be restructured such that edges do not overlap. For example, if two Twitter users interact on day 1 (duration: day 1 - day 5) and day 2 (duration: day 2 - day 6), the duration of these two interactions overlap and provide different information about the interaction between these two nodes on day 2 through day 5. To resolve this, we chose to truncate the first interaction: the duration of the first interaction becomes day 1 - day 1 and the duration of the second interaction remains day 2 – day 6. In this way, each edge is characterized by the most recent interaction between the two nodes that it connects.

To generate a series of network layouts for each timepoint, the compute.animation() function was applied to a networkDynamic object. The animation can be viewed in a web browser using the render.d3move() function. The resulting animation transitions through the network layouts for each timepoint showing how the network evolved over time. The color, shape, and size of nodes and edges can be characterized by their respective attributes, and tooltips can be utilized to display additional information about nodes and edges when clicked.

### 3.4.2 Path Visualizations

In dynamic Twitter networks, it is often interesting to explore the effect that an individual has on the network. Path visualizations show how the influence of an actor can spread through a network over time. The tPath() function from the tsna package can be used to identify the paths that are forward or backward reachable from a chosen actor in the network.

The transmissionTimeline() function from the ndtv package produces a particularly useful visualization for paths which displays the shortest path from the chosen actor to all users that are reachable by that actor plotted by time and generation. The generation represents the degree of separation from a user to the chosen actor. For example, generation one includes all users that interact directly with the chosen actor, while generation two includes all users who interact with users that have previously interacted with the chosen actor.

This path information can also be incorporated into dynamic visualizations as a node attribute. For example, nodes can change to a particular color or shape to indicate when they have entered the path forward form an individual. This can be a useful way to visualize the spread of an individual's influence through the network.

**4. Case Study: AI Leadership Contest: Musk vs. OpenAI Team**

The accelerating development and deployment of AI, particularly AI systems with capabilities surpassing current state-of-the-art models, presents an interesting case study for SNA. These tools offer ways to understand and monitor the societal implications and public perceptions surrounding AI evolution. The concerns about loss of control over AI systems and organizational risks, the potential spread of AI-generated misinformation and malicious use, job displacement due to automation and AI Race, and most profoundly "existential threats" to human civilization, all underscore the relevancy of this case study (Hendrycks et al., 2023; Future of Life, 2023). In an era where AI systems are increasingly integrated into our social fabric, understanding the dynamics of societal discourse, sentiment, and network structures related to AI aims to provide valuable insights into how these technologies are perceived, discussed, and potentially mitigated. It also allows us to observe, in real-time, the unfolding societal dialogue on AI's risks and benefits. This case study, therefore, offers a unique opportunity to apply social network and sentiment analysis in a context of high societal and academic relevance, contributing significantly to our understanding of AI's societal impact.

The three-month discussion between researchers is included in this study to structure a pathway for researchers having similar research questions. The following stages were identified throughout these discussions as a path to find the methodology;

a. **Identifying Supporters and Detractors of ChatGPT:** We initiated the process by identifying individuals who support or oppose ChatGPT. The language used by these groups, their distinct characteristics, their sentiment, and the influential figures amongst them were all of interest.
b. **Data Pulling Strategy:** Due to the enormity of the data corpus, we devised a targeted data pulling strategy. The focus was primarily on current news and topics within our domain.
c. **The open letter to stop training of AI:** To find the most influential individuals and observe the sentiment trajectory, we first pulled tweets from the date Elon Musk's signed letter was publicized in the main media (28th March 2023) and tracked backward until November 30[th], 2022, the date ChatGPT was released to the public, based on the keyword "ChatGPT".
d. **Sentiment Analysis and Plotting Sentiments by Day:** Various sentiment analysis packages like NLTK, TextBlob, VADER, DeepMoji, Transformers by Hugging Face, and Flair were tested to find the most suitable one for our study. Then we decided to plot the average sentiment by users, constrained by the keywords given above, per day to find the extreme changes to identify the events that happened in that time to see the influence and compare the impact of the letter.

e. **Dynamic Attributes:** To visualize the predominant sentiment and how it changed over time, we conceptualized the framework based on dynamic attributes that could be added to evaluate different features of text such as sentiment, length, and word count apart from the frequency of users' tweets.
f. **Keyword Coding:** By looking at the network based on the letter, the sentiment analysis results did not capture the change in the discourse that we were able to observe, therefore we tried to capture this change with the use of keyword coding based on the conversation.
g. **Forward Path:** Given two prominent figures in this conversation, Elon Musk and Sam Altman's influence, we analyzed their forward path and how they evolved over time concerning OpenAI tools.

In the methodology of this study, we pursued a comprehensive approach to understanding the influence of a letter calling for a pause in AI training on the discourse surrounding OpenAI's large language models (LLMs) and ChatGPT on X.com. The release of this letter marked the beginning of a significant public dispute between Elon Musk and Sam Altman, which centered around differing visions for AI's future: Musk advocating for stringent regulation to mitigate existential risks, and Altman favoring a transparent, cooperative development approach (Guardian, 2024). These contrasting responses to the letter exemplify the fundamental disagreements on governance and ethical direction, raising essential questions about the balance between AI innovation and public accountability. Our analysis aims to explore how this pivotal event influenced societal discussions and perceptions of AI, shaping the narrative in online and offline communities.

To comprehensively analyze the impact of the open letter's dissemination through main broadcasts on March 28, 2023, we employed a strategic data collection approach. We pulled tweets two-weeks prior to (3-15-2023) the letter and until two weeks after the letter (4-12-2023). The textual data of the tweets were coded both for sentiment and keywords to capture the changes in the conversation around the letter in network. Sentiment analysis utilized Python's TextBlob and VADER packages. We used the most salient words from the letter, which we determined were "risk", "danger" and "harm" to evaluate the progress of the conversation in terms of keywords. Furthermore, we created social networks, visualized conversations dynamically, and plotted path visualizations to reveal the interconnectedness and flow of the conversation across the platform. The following outlines our steps in the case study;

a. **Pulling Data:** To target knowledgeable users of ChatGPT, we pulled tweets about ChatGPT and large language models (LLM) using the following queries.
   a. OpenAI Generative
   b. OpenAI LLM
   c. OpenAI Large Language Models
   d. ChatGPT Generative
   e. ChatGPT LLM
   f. ChatGPT Large Language Models

b. **Creating Social Networks:** Mentions, replies, and retweets are used to form the edges of our network. We also pulled the profile information to produce an extensive nodelist. Then, some of the important people in the network are grouped into two teams, called: "Musk" and "OpenAI".
c. **Dynamic Visualization of the Conversation and Evolution of Sentiment and Keyword Use:** The text of the tweets were coded for sentiment and keyword use. The sentiment and keyword coding are aggregated to produce a measure for each individual at each time point they are active in the network. The sentiment and keyword measures for each individual are used to color code the nodes in the dynamic visualizations, which allowed us to easily observe changes over time. Node shapes are determined by team membership: Musk team is triangles, OpenAI team is squares, and those with no affiliation are circles. Node sizes are determined by the number of interactions at a given time point, the size of the node for a particular individual is allowed to change such that it becomes larger when they are more active and shrinks when they are less active. Due to the large size of the network, nodes with a total degree less than 10 are filtered out of the network visualizations.
d. **Path Visualizations:** In the case study, we tracked Elon Musk and Sam Altman's interactions over time with path visualizations. The forward paths of these two actors were generated separately so that we could compare their relative influence within this network.

**Table 1.** Analysis Table of Sentiment Scores Across Time Intervals: Two Examples.

| Onset | Terminus | From | To | Text | Sentiment (TextBlob) | Sentiment (VADER) |
|---|---|---|---|---|---|---|
| 17 | 21 | BigDataQueen_me | OpenAI | Just as I predicted last summer, authorities have started restricting young people's access to Large Language Models. Italy has banned #ChatGPT due to "privacy concerns," with one of the reasons being that @OpenAI doesn't verify the age of its users. | 0.047 | -0.681 |
| 22 | 26 | mStarJP | OpenAI | @intelligentHQ @OpenAI @Google @Microsoft @nvidia It's exciting to see how Generative AI like GPT-4 is improving its performance in language tasks, like passing simulated bar exams. Can't wait to see what other breakthroughs are on the horizon! | 0.072 | 0.883 |

The sentiment analysis, using both TextBlob and VADER methods before and after the letter's release, showed no major differences, as depicted in Figure 2. Both models generally pointed to a slightly positive sentiment, with the average just above zero which is the neutral value. The average daily sentiment value ranges from 0.020 to 0.181 using the VADER sentiment coding and 0.046 to 0.154 using the TextBlob sentiment coding, these variations are negligible considering both of these polarity scores have a full range from –1 to 1. After conducting evaluations using various test sentences, as exemplified in Table 1, we determined that the VADER algorithm exhibited better alignment with our judgement of sentiment. Consequently, we opted to utilize VADER for sentiment assessment in subsequent network visualizations.

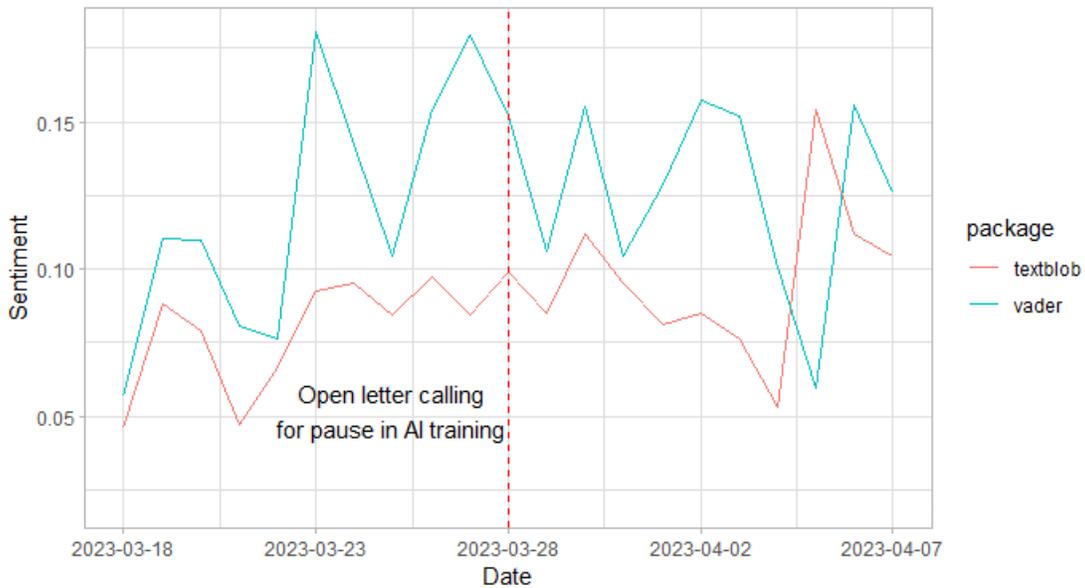

**Figure 2.** VADER and TextBlob Sentiment Analysis Change Before and After the Letter.

For our keyword coding, we focused on keywords used in Twitter interactions such as "risk," "danger," and "harm," inspired by the letter advocating caution in AI training. We monitored interactions involving these keywords over defined time intervals. Figure 3 presents the temporal trend of the keywords mentioned. Notably, on March 28, 2023, there was a pronounced increase in tweets containing these keywords. We found that a significant proportion of the spike in mentions was linked to retweets of a few influential accounts, amplifying the message within a short time frame. However, this increase was short-lived; by the subsequent day, the frequency returned to patterns similar to what was observed before the release of the letter, highlighting the transient nature of Twitter conversations. This rapid decline post-peak further underscores the importance of real-time monitoring to capture such fleeting yet impactful conversations in the vast landscape of social media.

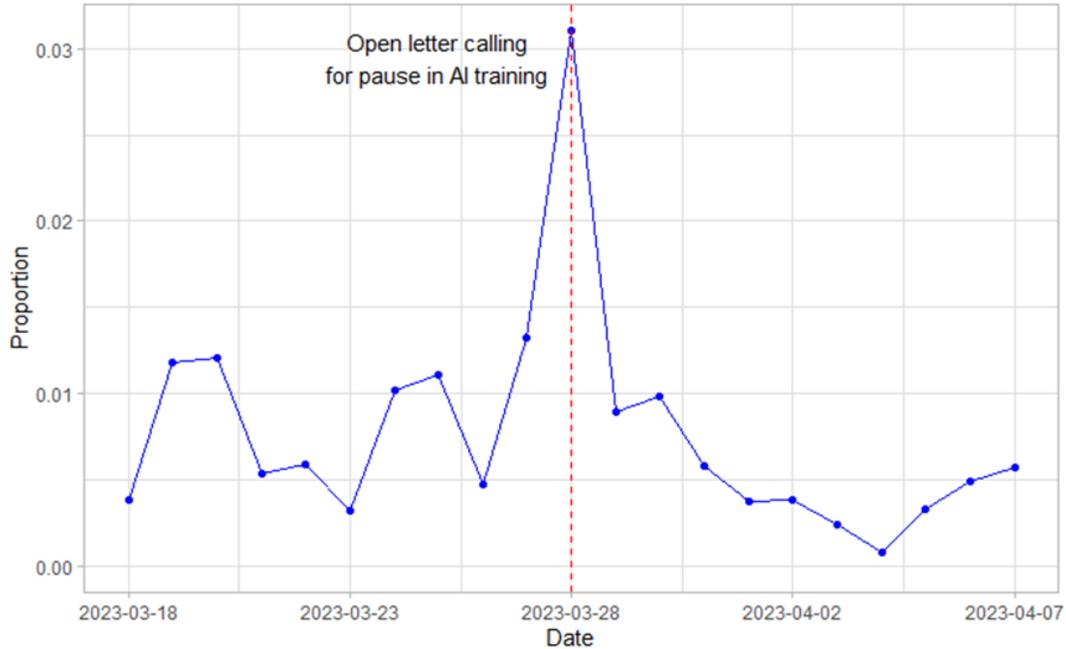

**Figure 3.** Keyword Analysis Results Throughout the Timeframe in the Network.

Our customized dynamic visualizations were designed to provide a more comprehensive understanding of the profile interactions and their progression over time. In the context of our study, positive sentiment was denoted by green-colored nodes, neutral sentiment by yellow, and negative sentiment by red, using the VADER sentiment algorithm in the upper three networks. The red color represents the use of the selected keywords in the lower three networks. The dataset spans across 29 temporal slices, accounting for both preceding and subsequent days surrounding the release of the open letter. Sentiment and keyword analyses are illustrated for March 23, March 28, and April 2. Network graphs positioned directly below one another correspond to sentiment and keyword analyses for the same date. As depicted in the following figure, distinct shapes were assigned to different groups; the Musk group was represented as triangles, the OpenAI group as squares, and all others are circles. Within the Figure, node sizes were determined based on the degree on each date, scaled from 0.3 to 1.3, corresponding to counts that ranged from 1 to 2879. In the animations generated by the implemented code, the nodes and edges can be clicked to display profile and tweet information, respectively. Consequently, capturing and exploring the dynamic nature of the network becomes a straightforward task. Overall, 1323 nodes were represented in the network, however, the graphs below only represent the portion of nodes that were active at each given date.

Sentiment scores presented are based on VADER analysis and were observed to range from a minimum of -0.802 to a maximum of 0.950. An exemplar of the minimum sentiment is observed in a tweet warning against malevolent use of AI technologies, while a positively tinted tweet highlights the rise and utility of ChatGPT for educational purposes.

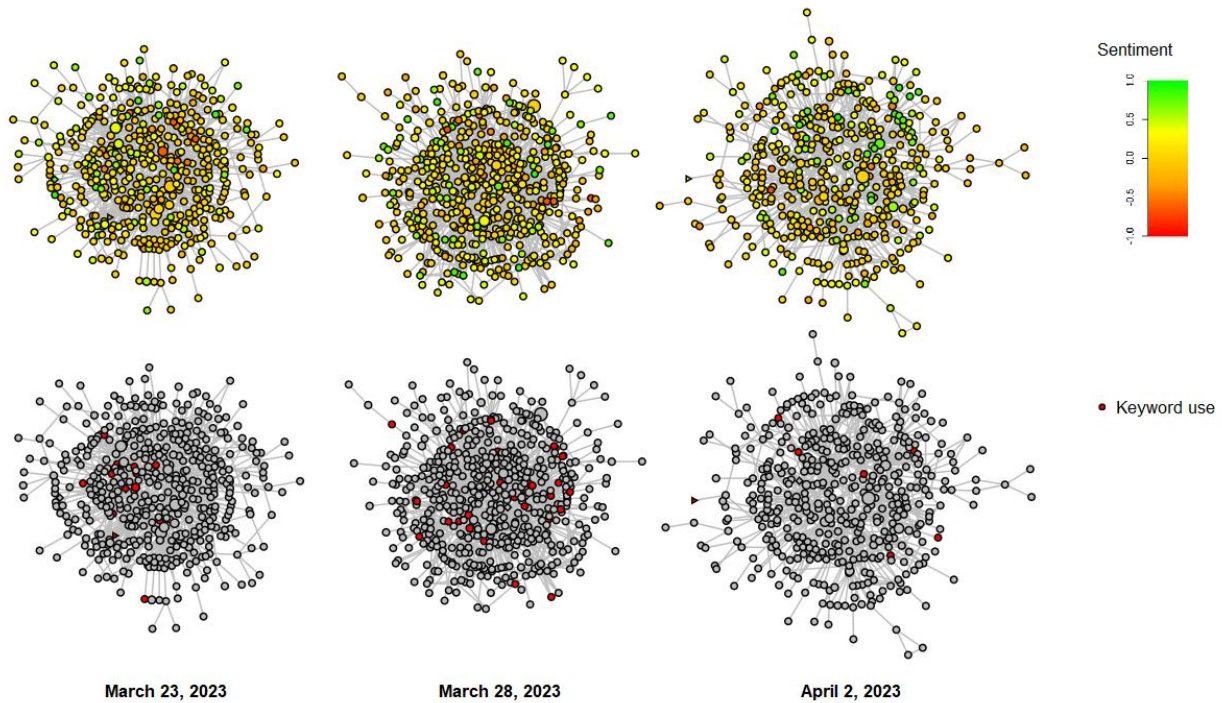

**Figure 4.** Sentiment Analysis (upper three) and Keyword Coding (lower three) Results.

In our study, sentiment analysis was employed to interpret Twitter user sentiments both before and after the broadcasting of the letter. The results indicated a marginal variation in overall sentiment. The network visuals make differences more discernible based on the representation and sizes of clustered red (negative sentiment) and green (positive sentiment) nodes. On March 23, before the public broadcast of the letter, a small clustering of red nodes was visible, centering around one larger-sized red node, indicating influential accounts sharing negative views on AI training, which could be attributed to the initial publication on March 22, 2023 (Future of Life, 2023). By March 28, sentiment appeared more neutral, with a decline in larger nodes, suggesting fewer dominant and skewed opinion leaders. On April 2, the landscape had shifted slightly again, with a small cluster of green nodes, hinting at a somewhat more positive conversation, possibly due to influential users coming to the defense of ChatGPT against the claims stated in the letter. Overall, the variations in sentiment over time were minimal, indicating that this letter had very little influence on sentiment.

The keyword analysis captured a more meaningful effect of the letter. The increased presence of red nodes on the day the letter was broadcasted, supported by Figure 4, demonstrates the utility of this type of network visualization. Keyword usage manifested in a more dispersed pattern, rather than clusters, hinting at a broader linguistic evolution across the network instead of a few influential tweets with a high number of interactions. As such, the keyword analysis became important to highlight the network's dynamic linguistic transitions.

Integrating both the sentiment and keyword analyses, it was observed that the OpenAI team (symbolized as squares) participated in discussions utilizing the identified keywords both prior to and following the letter's release. Figure 5 highlights one such interaction and demonstrates how tooltips can be used to display edge information in the dynamic visualizations. The sentiment of these interactions consistently ranged from neutral to mildly positive. This measured approach might have been deliberate, as stronger advocations could potentially raise suspicions among the public. Such a strategy could be aimed at effectively managing the situation and perceptions.

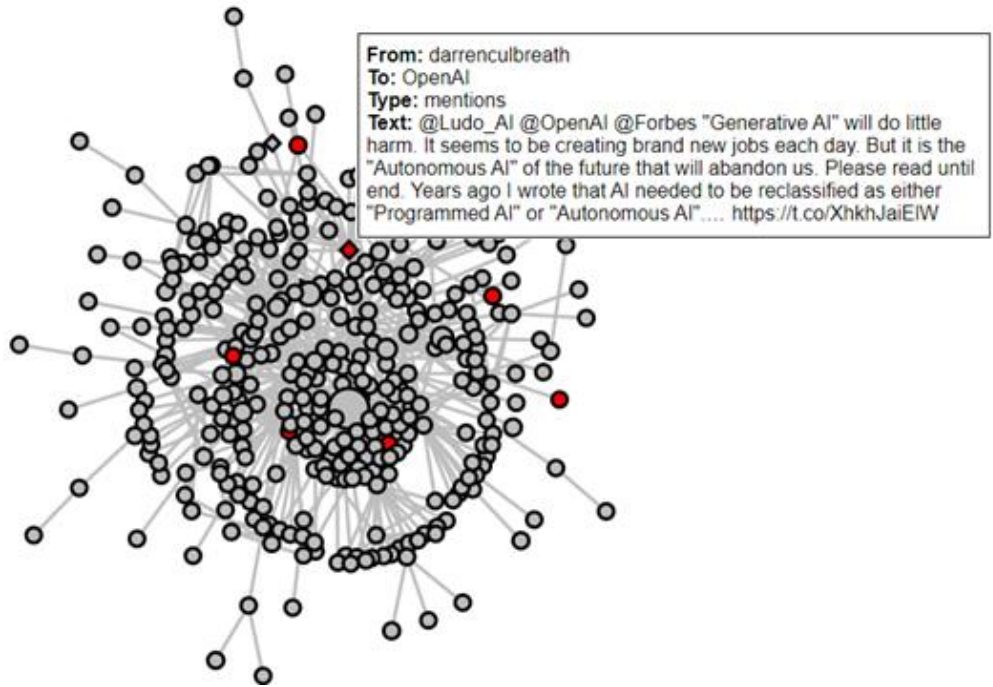

**Figure 5.** Tooltip display of edge characteristics (March 19, 2023).

Elon Musk did not directly tweet about the letter, even though he had endorsed it. In our study, we observed a notable trend concerning interactions with Elon Musk across the network. Specifically, the frequency of interactions was highest near the end of the period under investigation, starting from our designated initial date. These interactions predominantly exhibited a neutral tone. The "degree" column in the accompanying table quantifies the peak number of interactions for each respective interval, while the "sentiment" column provides the average sentiment score for the same time frame.

In the nodelist, dynamic attributes are updated for any day in which a new interaction occurs. In the network, each edge persists for a 5-day period, and text analysis variables are updated with each new interaction. The choice of a 5-day duration aims to strike a balance; shorter intervals resulted in visuals that were less informative and underwent rapid, unhelpful changes, thereby not providing an optimal representation. A new interaction is characterized as a fresh retweet, reply, or mention involving a specific individual, and these time intervals are anchored by such new interactions. For instance, if a user engages in an interaction on Day 1, the corresponding node attributes will be retained until Day 5. If another interaction occurs before Day 5, the text analysis variables are adjusted to account for these new interactions and appear as a new row in the nodelist. Table 2 provides a comparative analysis of interactions and sentiment values, specifically showcasing patterns associated with Elon Musk and Sam Altman over various intervals.

**Table 2.** Comparative Interaction and Sentiment Analysis Table.

| Elon Musk | | | | Sam Altman | | | |
|---|---|---|---|---|---|---|---|
| Onset (Day) | Terminus (Day) | Degree of Interactions | Sentiment (VADER) | Onset (Day) | Terminus (Day) | Degree of Interactions | Sentiment (VADER) |
| 1 | 4 | 17 | 0.213 | 1 | 2 | 13 | 0.351 |
| 4 | 7 | 9 | 0.089 | 2 | 3 | 13 | 0.432 |
| 7 | 10 | 16 | 0.052 | 3 | 4 | 9 | 0.324 |
| 10 | 11 | 16 | -0.101 | 4 | 7 | 9 | 0.229 |
| 11 | 12 | 16 | 0.168 | 7 | 10 | 5 | 0.358 |
| 12 | 15 | 11 | -0.012 | 10 | 12 | 17 | 0.096 |
| 15 | 17 | 9 | -0.046 | 12 | 13 | 17 | 0.116 |
| 17 | 21 | 8 | 0.202 | 13 | 16 | 17 | 0.328 |
| 22 | 26 | 19 | -0.05 | 16 | 17 | 9 | 0.358 |
| - | - | - | - | 17 | 18 | 9 | 0.253 |
| - | - | - | - | 18 | 19 | 9 | 0.12 |
| - | - | - | - | 19 | 22 | 8 | 0.427 |
| - | - | - | - | 22 | 23 | 8 | 0.483 |
| - | - | - | - | 23 | 25 | 8 | 0.281 |
| - | - | - | - | 25 | 27 | 213 | 0.125 |
| - | - | - | - | 27 | 28 | 213 | 0.062 |
| - | - | - | - | 28 | 32 | 213 | 0.031 |

Figure 6 represents the forward paths of Elon Musk and Sam Altman throughout this conversation. The timelines demonstrate the spread of the influence of these individuals through this network. From these visual representations, it is apparent that Elon Musk's influence spread a few days sooner than Sam Altman's. Overall, the number of forward-reachable users was similar, with Elon Musk reaching 799 individuals and Sam Altman reaching 708.

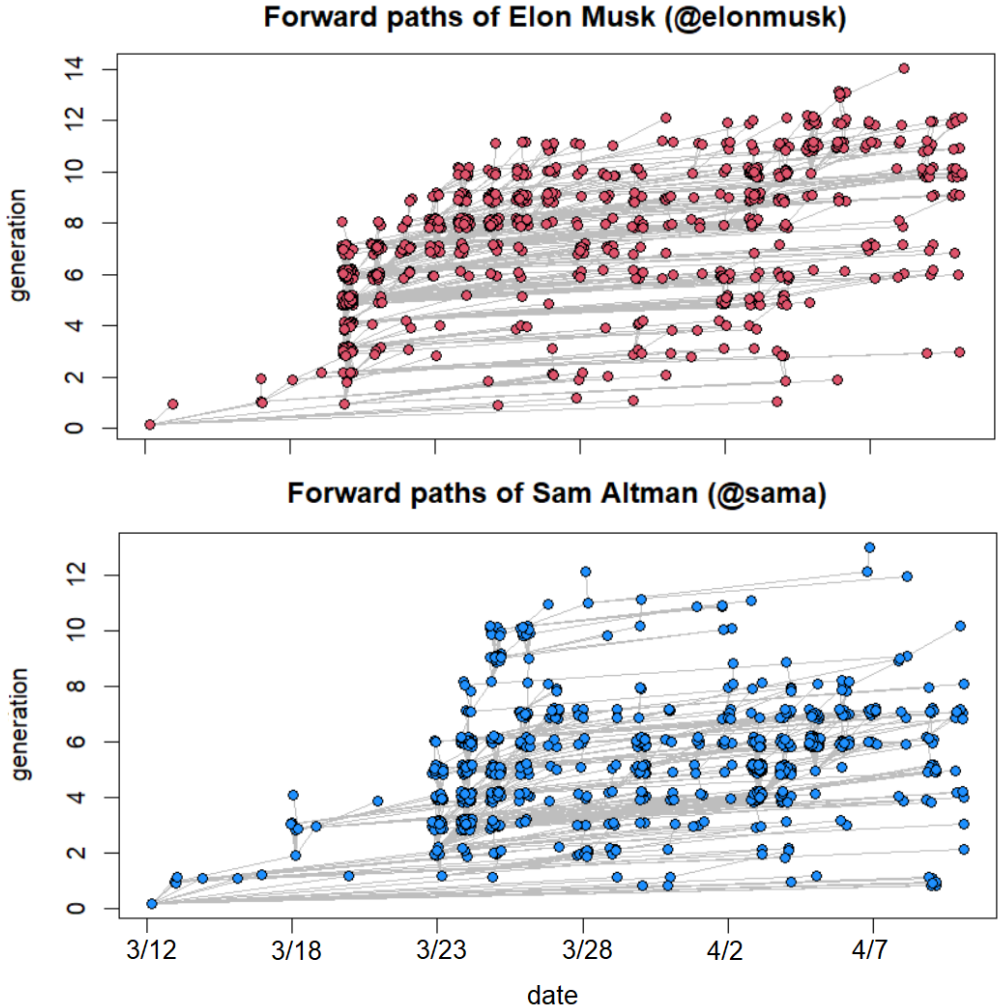

**Figure 6.** Forward Path Results.

Figure 7 illustrates the evolution of Elon Musk's and Sam Altman's influence networks over time. The red nodes represent individuals included in Elon Musk's forward path, blue nodes represent those in Sam Altman's forward path, and purple nodes represent individuals who are included in both forward paths. This visual representation highlights the overlap in the two forward paths. Most individuals in the network are forward reachable by both actors. Interestingly, there are very few blue nodes in the network, indicating that almost all individuals who are in Sam Altman's path of influence were in Elon Musk's forward path first. By examining these changes over time, we can better understand how conversations evolve and whether specific individuals have been influenced by one, both, or neither of the figures in focus during the specified time frame.

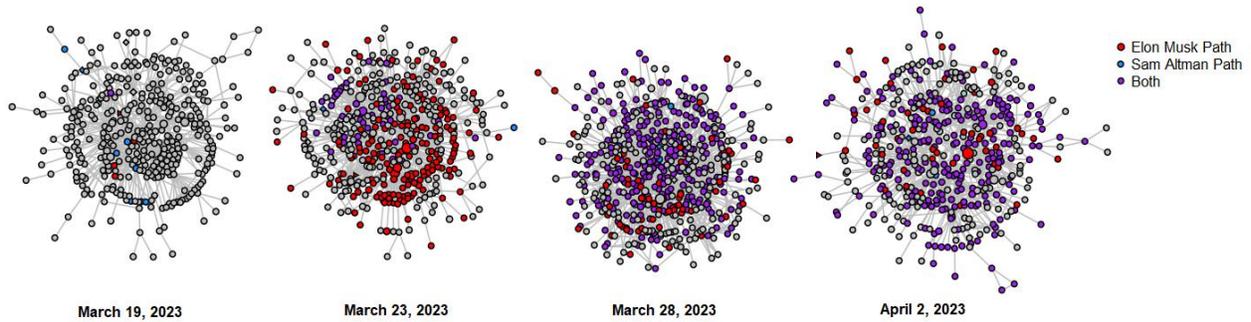

**Figure 7.** Forward Path Networks of Elon Musk and Sam Altman (March 19 to April 2, 2023).

Notice on March 19 that the majority of the network is not contained in either forward path. This does not mean that none of these individuals interacted with Elon Musk or Sam Altman before this date, only that they did not interact with them since the beginning of our observation period, which began on March 15. We examined the same three dates as our sentiment and keyword comparison. During this period, from March 23 to March 28, there's a significant shift of individuals into Sam Altman's path of influence. Triangles, representing Musk Team and rectangles representing OpenAI team could be followed in the figures to follow the interactions. Figure 8 shows a closer look at how this change happened in this network.

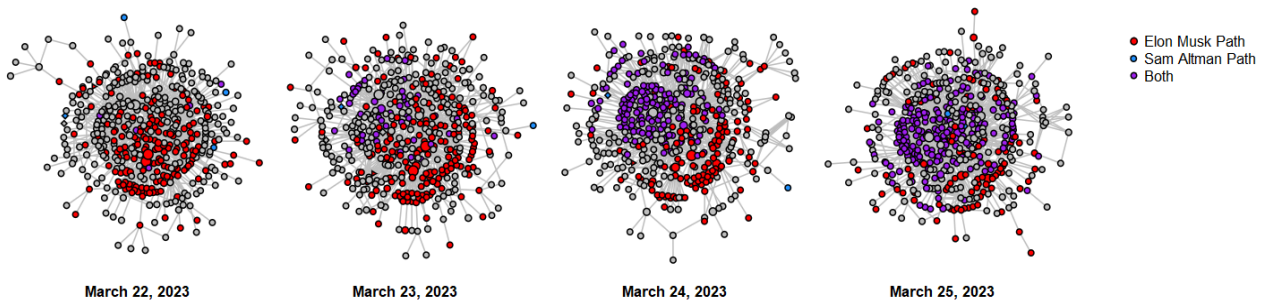

**Figure 8.** Forward Path Networks of Elon Musk and Sam Altman (March 22 to March 25, 2023).

The spread of Sam Altman's influence in this network happened rapidly. Up until March 22, there were many Twitter users in Elon Musk's forward path but very few in Sam Altman's. On March 23, we begin to see a noticeable cluster of individuals who are in both paths of influence and by March 25, the majority of the network is in both forward paths. This all happens before the public release of the letter on March 28, so the changes in forward paths in this network appear to be unrelated to the letter.

## 5. Results and Discussion

In today's digital era, understanding the dynamics of social media discourse is of utmost importance. Our study, while centered on the discourse around the open letter on AI training, primarily introduced a methodology for analyzing such discussions. Using a combination of Python and R packages, enhanced with custom coding, we developed tools to effectively extract, process, and analyze Twitter data for social network research after comprehensive literature research.

Our method is structured in three stages. First, we used R to pull Twitter data and generate social network datasets. Second, we integrated Python and R code to conduct sentiment analysis and keyword coding. This allowed us to detect shifts in sentiment and language use regarding our chosen topic. Lastly, we utilized R packages for dynamic and path visualizations, giving a clear representation of evolving social networks. Using the debate around ChatGPT as a case study, we demonstrated the efficacy of our approach. It was particularly effective in capturing changes in sentiment and language on March 28, 2023. The visualizations highlighted the changing nature of user connections in the debate and emphasized the key roles of influential figures.

We saw changes in the use of specific language in the network based on our keyword analysis, however the overall sentiment stayed the same over time. We also looked at forward paths and combined them with dynamic network visualizations to see how certain individuals influenced the conversation.

However, our study faced several limitations. The recent changes to Twitter's API, combined with lack of academic access, severely restricted our ability to gather sufficient data, limiting our analysis to just one month. We focused only on the forward paths of two key figures due to their relevance. While our primary focus was on visualization methodologies, this leaves several aspects of social network analysis unexplored for future research, such as influence and selection modeling, or dynamic modeling. In conclusion, our methodology provides a comprehensive approach to understanding social media discourse on platforms like X.com, laying a solid groundwork for future studies.

**Author's Contribution Statement**

All authors collaboratively conceptualized, designed, and implemented this study. Both authors actively participated in writing and iteratively developing the manuscript. The first author primarily prepared the code, refining it through discussions with the co-author, which also helped enhance the manuscript at various stages. Both authors crafted each draft, integrating their insights and expertise. The final manuscript was jointly reviewed and approved by all authors.

**Financial Support and Conflict of Interest Statement**

This research study did not receive any external funding, and authors did not declare any competing interest.

**Data Availability Statement**

All data supporting the findings of this study, including methodology visuals, are included and accessible at the GitHub repository: https://github.com/nrjess/twitter-sna. This repository provides access to the datasets used in research, along with the relevant visuals that illustrate methodology.